\documentclass{aa}
\usepackage[varg]{txfonts}
\usepackage{graphicx}
\usepackage{ulem}
\usepackage{color}

\usepackage{natbib}
\bibpunct{(}{)}{;}{a}{}{,} 

\definecolor{orange}{rgb}{ 0.95, 0.60, 0}

\begin{document}

\title{Formation of terrestrial planets in disks evolving via disk winds and implications for the origin of the solar system's terrestrial planets}
\subtitle{}
\titlerunning{Formation of terrestrial planets in disks evolving via disk winds}

\author{Masahiro Ogihara\inst{1}
\and Hiroshi Kobayashi\inst{2}
\and Shu-ichiro Inutsuka\inst{2}
\and Takeru K. Suzuki\inst{2}
}
\institute{Observatoire de la C\^ote d'Azur, Boulevard de l'Observatoire, 06304 Nice Cedex 4, France \email{omasahiro@oca.eu}
\and Nagoya University, Furo-cho, Chikusa-ku, Nagoya, Aichi 464-8602, Japan}
\date{Received 9 January 2015 / Accepted 20 April 2015}

\abstract 
{
Recent three-dimensional magnetohydrodynamical simulations have identified a disk wind by which gas materials are lost from the surface of a protoplanetary disk, which can significantly alter the evolution of the inner disk and the formation of terrestrial planets. A simultaneous description of the realistic evolution  of the gaseous and solid components in a disk may provide a clue for solving the problem of the mass concentration of the terrestrial planets in the solar system.
} 
{
We simulate the formation of terrestrial planets from planetary embryos in a disk that evolves via magnetorotational instability and a disk wind. The aim is to examine the effects of a disk wind on the orbital evolution and final configuration of planetary systems.
} 
{
We perform \textit{N}-body simulations of sixty 0.1 Earth-mass embryos in an evolving disk. The evolution of the gas surface density of the disk is tracked by solving a one-dimensional diffusion equation with a sink term that accounts for the disk wind. 
}
{
We find that even in the case of a weak disk wind, the radial slope of the gas surface density of the inner disk becomes shallower, which slows or halts the type I migration of embryos. If the effect of the disk wind is strong, the disk profile is significantly altered (e.g., positive surface density gradient, inside-out evacuation), leading to outward migration of embryos inside $\sim 1 {\rm AU}$.
} 
{
Disk winds play an essential role in terrestrial planet formation inside a few AU by changing the disk profile. In addition, embryos can undergo convergent migration to $\sim 1 {\rm AU}$ in certainly probable conditions. In such a case, the characteristic features of the solar system's terrestrial planets (e.g., mass concentration around 1 AU, late giant impact) may be reproduced.
}
\maketitle

\section{Introduction}
\label{sec:intro}
Planets with masses larger than approximately $0.1 M_\oplus$ significantly excite density waves in a protoplanetary disk, which in turn exerts torque on the planets, and migrate toward the central star (type I migration) (e.g., \citealt{goldreich_tremaine80}; \citealt{ward86}). It is known that the inward migration timescale for an Earth-mass planet at 1 AU is $\sim 10^5$ years in a locally isothermal disk (e.g., \citealt{tanaka_etal02}), which is shorter than the disk lifetime and thus can be a problem for terrestrial planet formation in the solar system. Recent studies have shown that in an adiabatic disk, fully unsaturated positive corotation torques can compensate for negative Lindblad torques, which can change the direction and rate of migration (e.g., \citealt{paardekooper_etal10}). If this effect is significant, a convergence zone where migration is convergent can be created in a disk and gas giant cores can form at a few tens of AU \citep{horn_etal12}.

\citet{paardekooper_etal11} revised the torque formula by taking into account the effects of viscous and thermal diffusion on the corotation torque, which indicates that only planets with a limited range of masses can experience the non-linear corotation torque due to saturation effects. In fact, planets with masses smaller than a few Earth masses are not affected by the non-linear corotation torque (the horseshoe torque) and they migrate inward under the influence of the Lindblad torque (e.g., \citealt{kretke_lin12}; \citealt{hellary_nelson12}; \citealt{cossou_etal14}), which means that the solar system's terrestrial planets still have the problem of inward migration. However, if the slope of the gas surface density becomes large enough, the positive linear corotation torque can reverse the migration.

During the evolution of protoplanetary disks, vertical winds as well as radial accretion play a significant role in the dispersal of the gas component. While photoevaporating winds by UV and X-rays from a central star have been widely discussed (e.g., \citealt{alexander_etal06}; \citealt{ercolano_etal09}), the importance of spontaneously-driven disk winds has also been pointed out (\citealt{suzuki_inutsuka09}; \citealt{suzuki_etal10});  using magnetohydrodynamic (MHD) simulations they showed that MHD turbulence triggered by magnetorotational instability (MRI) in disks  inevitably drives vertical outflows as well as the radial transport of angular momentum and consequent accretion. The detailed properties of such disk winds driven by turbulent Poynting flux have been  studied \citep{bai_stone13,fromang_etal13,lesur_etal13}. An intriguing characteristic of the disk wind is that  disk dispersal proceeds gradually in an inside-out manner because the dispersal timescale is approximately proportional to rotation time $\propto r^{3/2},$ where $r$ is the radial distance from the star. As a result, the surface density gradient at $r\lesssim1 {\rm AU}$ becomes much shallower than the typical values of -0.5 to -1.5, and could even be positive, which will drastically affect the migration of protoplanets.

As we are interested in seeing how planet formation proceeds in such a disk, we investigate the formation of terrestrial planets using \textit{N}-body simulations in a disk where the effect of a disk wind is considered. We calculate the disk evolution using a one-dimensional disk with viscous diffusion and a disk wind based on \citet{suzuki_etal10}. In this work, we focus on the late phase of planet formation when the effect of a disk wind is clearly visible, and calculate the accretion and orbital evolution of planets from planetary embryos with masses of $0.1 M_\oplus$ for $\sim 100$ Myr. The \textit{N}-body code includes the effect of type I migration that depends on the slope of gas surface density and the mass of the planets using the prescription described by \citet{paardekooper_etal11}. We also include the influence of eccentricity on the corotation torque. 

Our primary aim is to clarify the effect of a disk wind on terrestrial planet formation. In addition, we also discuss the origin of the terrestrial planets in the solar system. A number of authors (e.g., \citealt{kominami_ida02}; \citealt{nagasawa_etal05}; \citealt{ogihara_etal07}; \citealt{raymond_etal09}; \citealt{morishima_etal10}) have studied the origin of these planets using \textit{N}-body simulations; however, no single simulation has yet convincingly reproduced the observed constraints simultaneously (e.g., eccentricities, spatial concentration around 1 AU, late moon-forming giant impact). We will propose a possible model for the origin of the inner solar system.

The plan of the paper is as follows. In Sect.~\ref{sec:model} we describe our model and the initial conditions of the simulations; in Sect.~\ref{sec:results} we present the results of \textit{N}-body simulations; in Sect.~\ref{sec:discussion} we discuss the origin of the terrestrial planets in the solar system; in Sect.~\ref{sec:conclusions} we give our conclusions.

\section{Model description}
\label{sec:model}
\subsection{Disk model}
\label{sec:disk_model}
We calculate the time evolution of the gas component of a protoplanetary disk by taking into account the radial mass flows by the transport of angular momentum through MHD turbulence and the mass loss due to the disk wind. Although there are uncertainties in the mass flux of the disk wind, we follow the simple framework of \citet{suzuki_etal10}. The evolution of gas surface density $\Sigma_{\rm g}$ can be expressed in the form of a diffusion equation,
\begin{eqnarray}
\label{eq:diffusion}
\frac{\partial \Sigma_{\rm g}}{\partial t} = \frac{3}{r} \frac{\partial}{\partial r} \left[r^{1/2} \frac{\partial}{\partial r} (\nu \Sigma_{\rm g} r^{1/2})\right] - C_{\rm w} \frac{\Sigma_{\rm g}}{\sqrt{2 \pi}} \Omega,
\end{eqnarray}
where $\Omega$ is the Keplerian frequency. The standard $\alpha$ viscosity prescription, $\nu = \alpha c_{\rm s } h,$ is used, where $c_{\rm s}$ indicates the sound velocity and $h$ is the disk scale height. The second term on the right-hand side of Eq.~(\ref{eq:diffusion}) is the disk wind flux $\rho v_z$, which can be expressed as $C_{\rm w} \rho_0 c_{\rm s}$ using the mid-plane density $\rho_0$, $c_{\rm s}$, and a non-dimensional constant $C_{\rm w}$ \citep{suzuki_etal10}. We note  that the coefficient of the first term on the right-hand side of Eq.~(\ref{eq:diffusion}) is not exactly the same as  that of Eq.~(9) of \citet{suzuki_etal10} owing to a different definition of $\nu$, which gives a slightly different evolution of the disk surface density.

For the initial distribution of the gas disk, we assume
\begin{eqnarray}
\label{eq:initial_disk}
\Sigma_{\rm g,ini} = 2400 f_{\rm g} \left(\frac{r}{1 \rm{AU}} \right)^{-3/2} \exp\left(-\frac{r}{r_{\rm cut}}\right)\,\mathrm{g\, cm}^{-2},
\end{eqnarray}
where $f_{\rm g}$ is a dimensionless parameter and $r_{\rm cut} = 50 {\rm AU}$ is used as a cutoff radius. In our $N-$body code, we use the evolution of $\Sigma_{\rm g}$ to calculate the effect of the gas disk on planet evolution.

The temperature distribution is assumed to be that of an optically thin disk \citep{hayashi81},
\begin{eqnarray}
T = 280 \left(\frac{r}{1 {\rm AU}} \right)^{-1/2} {\rm K}.
\end{eqnarray}
We note that although type I migration is also dependent on the temperature profile, we fix the temperature distribution to clarify the effect of the disk wind via surface density distribution. Then the disk scale height $h$ is given by
\begin{eqnarray}
h/r = 0.047 \left(\frac{r}{1{\rm AU}}\right)^{1/4}
\left(\frac{L_*}{L_\odot}\right)^{1/8}
\left(\frac{M_*}{M_\odot}\right)^{-1/2},
\end{eqnarray}
where $L_*$ and $M_*$ are the luminosity and mass of the host star, respectively.

\subsection{Type I migration and eccentricity damping}
The orbits of planetary embryos with masses $M_{1}, M_{2}, ...$ and position vectors $\textbf{\textit{r}}_1, \textbf{\textit{r}}_2, ...$ relative to the host star are calculated by numerically integrating the equation of motion,
\begin{eqnarray}
\frac{d^2 \textbf{\textit{r}}_k}{dt^2}
&=&  -GM_* \frac{\textbf{\textit{r}}_k}{ |\textbf{\textit{r}}_k|^3} 
- \sum_{j \neq k} GM_j 
\frac{\textbf{\textit{r}}_k - \textbf{\textit{r}}_j}{|\textbf{\textit{r}}_k - \textbf{\textit{r}}_j|^3} 
- \sum_{j} GM_j \frac{\textbf{\textit{r}}_j}{ |\textbf{\textit{r}}_j|^3}\nonumber\\
&&+ \textbf{\textit{F}}_{\rm damp} + \textbf{\textit{F}}_{\rm mig},
\end{eqnarray}
where $\textbf{\textit{F}}_{\rm damp}$ is a specific force for eccentricity and inclination damping and $\textbf{\textit{F}}_{\rm mig}$ is a specific force for type I migration (see \citet{ogihara_etal14} for each formula). The timescale for damping of the eccentricity, $t_e$, is given by
\begin{eqnarray}
\label{eq:e-damp}
t_e &=& \frac{1}{0.78}\left(\frac{M}{M_*}\right)^{-1} 
\left(\frac{\Sigma_{\rm g} r^2}{M_*}\right)^{-1}
\left(\frac{c_{\rm s}}{v_{\rm K}}\right)^{4} \Omega^{-1}\\
&\simeq& 3 \times 10^2  f_{\rm g}^{-1}
\left(\frac{r}{1 {\rm AU}}\right)^2
\left(\frac{M}{M_\oplus}\right)^{-1}
\left(\frac{M_*}{M_\odot}\right)^{-1/2}
\left(\frac{L_*}{L_\odot}\right)^{1/2}
{\rm ~yr},
\end{eqnarray}
where $\Sigma_{\rm g} = 2400 f_{\rm g} (r/1 {\rm AU})^{-3/2} \,\mathrm{g\, cm}^{-2}$ is used in the second equality. Here the relative motion between gas and planets is assumed to be subsonic ($e v_{\rm K} \lesssim c_{\rm s}$). For planets with high eccentricities and inclinations, we include a correction factor according to Eqs.~(11) and (12) of \citet{cresswell_nelson08}. 

The migration timescale, $t_a$, is given by
\begin{eqnarray}
\label{eq:a-damp}
t_a &=& \frac{1}{\beta} \left(\frac{M}{M_*}\right)^{-1}
\left(\frac{\Sigma_{\rm g} r^2}{M_*}\right)^{-1}
\left(\frac{c_{\rm s}}{v_{\rm K}}\right)^{2} \Omega^{-1}\\
&\simeq& 2 \times 10^5  f_{\rm g}^{-1} \beta^{-1}
\left(\frac{r}{1 {\rm AU}}\right)^{3/2}
\left(\frac{M}{M_\oplus}\right)^{-1}
\left(\frac{M_*}{M_\odot}\right)^{1/2}
\left(\frac{L_*}{L_\odot}\right)^{1/4}
{\rm ~yr},
\end{eqnarray}
where $\beta$ is a coefficient that determines the direction and speed of type I migration. According to \citet{paardekooper_etal10}, the type I migration torque depends on the Lindblad torque, the barotropic part of the horseshoe drag (or the non-linear corotation torque), the entropy-related part of the horseshoe drag, the barotropic part of the linear corotation torque, and the entropy-related part of the linear corotation torque. \citet{paardekooper_etal11} derived the total type I migration torque including both saturation and the cutoff at high viscosity. Thus we write the migration coefficient in the form
\begin{eqnarray}
\beta = \beta_{\rm L} + \beta_{\rm c,baro} + \beta_{\rm c,ent},
\end{eqnarray}
where $\beta_{\rm L},$ $\beta_{\rm c,baro},$ and $\beta_{\rm c,ent}$ are related to the Lindblad torque, the barotropic part of the corotation torque, and the entropy-related part of the corotation torque, respectively. Each formula is given as 
\begin{eqnarray}
\label{eq:beta_lin}
\beta_{\rm L} &=& \frac{2}{\gamma} (-2.5 - 1.7q + 0.1p),\\
\beta_{\rm c,baro} &=& \frac{2}{\gamma} \Biggl(F(P_\nu) G(P_\nu) 1.1 \left[\frac{3}{2} - p \right] 
\nonumber \\
&&+ \left[ 1- K(P_\nu)\right] 0.7 \left[\frac{3}{2} - p \right] \Biggr),\\
\beta_{\rm c,ent} &=& \frac{2}{\gamma} \Biggl(F(P_\nu) F(P_\chi) \sqrt{G(P_\nu) G(P_\chi)} 7.9 \frac{\xi}{\gamma}\nonumber\\
&&+ \sqrt{(1- K(P_\nu)) (1- K(P_\chi))} \left[2.2 - \frac{1.4}{\gamma} \right] \xi \Biggr),
\end{eqnarray}
where $-p$ and $-q$ denote the local surface density gradient $(p(r) = -d \ln \Sigma_{\rm g}/d \ln r)$ and the local temperature gradient ($q(r) = -d \ln T/d \ln r$), $\gamma$ is the adiabatic index, and $-\xi$ is the local entropy gradient $(\xi = q - (\gamma - 1) p)$. In this work, $\gamma = 1.4$ and $q = 0.5$. The function $F(P)$ is a decreasing function with the value of [0,1] that is related to saturation, and the functions $G(P)$ and $K(P)$ are increasing functions with the value of [0,1] related to cutoff. \citet{paardekooper_etal11} introduced viscous and thermal parameters $P_\nu$ and $P_\chi$, which are expressed by the ratio between the  viscous/thermal diffusion timescales  $\tau_\nu/\tau_\chi$ and the horseshoe libration timescale $\tau_{\rm lib}$,
\begin{eqnarray}
P_\nu = \frac{2}{3} \sqrt{\frac{\Omega r^2 x_s^3}{2 \pi \nu}} \left(= \sqrt{\frac{16}{27} \frac{\tau_\nu}{\tau_{\rm lib}}}\right),\\
P_\chi =  \sqrt{\frac{\Omega r^2 x_s^3}{2 \pi \chi}}  \left(= \sqrt{\frac{4}{3} \frac{\tau_\chi}{\tau_{\rm lib}}}\right),
\end{eqnarray}
where the dimensionless half-width of the horseshoe region is
\begin{eqnarray}
x_s = \frac{w_s}{r} = \frac{1.1}{\gamma^{1/4}} \sqrt{\frac{M}{M_*} \frac{r}{h}},
\end{eqnarray}
where $w_s$ is the half-width of the horseshoe region. For prescriptions of $F(P)$, $G(P)$, and $K(P)$, we refer the readers to Eqs.~(23), (30), and (31) in \citet{paardekooper_etal11}. 

For the thermal diffusivity, we assume $\chi = \nu$ in this work for simplicity. The thermal diffusivity determines the saturation of the entropy-related part of the corotation torque. The thermal diffusivity is considered to be large when the disk surface density decreases and the disk is optically thin, and thus radiative cooling is efficient. In this case, the entropy-related corotation torque diminishes, and one might speculate that the orbital evolution of planets would change. However, we do not expect that this can change our results in this paper. In the terrestrial planet forming region, the disk opacity is about $1-10\, {\rm cm^2\, g^{-1}}$ (e.g., \citealt{bell_lin94}). This means that the disk is optically thin only when the disk is significantly dispersed $(\Sigma_{\rm g} \sim 0.1-1\, {\rm g\, cm^{-2}})$ and type I migration is no longer effective. Therefore, our simplified model $(\chi=\nu)$ may be justified. In addition, we observe that our qualitative results are not affected in test simulations in which the entropy-related torque is switched off from the beginning of the simulations.

Recent studies (e.g., \citealt{bitsch_kley10}) suggest that the corotation torque decreases as the planet eccentricity increases; therefore, we also consider this effect using the  formulae \citep{fendyke_nelson14}, 
\begin{eqnarray}
\beta_{\rm C,baro}(e) &=& \beta_{\rm C,baro} \exp\left(-\frac{e}{e_{\rm f}} \right),\\
\beta_{\rm C,ent}(e) &=& \beta_{\rm C,ent} \exp\left(-\frac{e}{e_{\rm f}} \right),
\end{eqnarray}
where $e_{\rm f} = 0.5h/r + 0.01$. We note that the Lindblad torque can also be reduced when planets have high eccentricities and inclinations (e.g., Papaloizou \& Larwood 2000). In some runs of our simulations, we also add a reduction factor to Eq.~(\ref{eq:beta_lin}), which is given by Eq.~(13) of \citet{cresswell_nelson08}. However, we found that even if this factor is ignored, the results do not change very much.

There are several works that take into account the effect of magnetic field on type I migration (\citealt{terquem03}; \citealt{fromang_etal05}; \citealt{muto_etal08}; \citealt{uribe_etal15}; \citealt{bans_etal15}). Among these works, \citet{bans_etal15} consider the vertical transport of angular momentum by disk winds. In our paper, we assume the net vertical magnetic field is very weak and the disk wind is driven by the MRI triggered turbulent magnetic field, which is in contrast to the disk wind by relatively strong net magnetic field assumed in \citet{bans_etal15}. Therefore, the type I migration formula based on pure hydrodynamics (which corresponds to the limit of weak magnetic field) by \citet{paardekooper_etal11} is probably still reasonable in our setting.

\subsection{Set of parameters and orbital integration}
\label{sec:parameters}
In our series of simulations, we mainly vary the following parameters: turbulent viscosity $\alpha$, disk wind efficiency $C_{\rm w}$. Although the efficiency of the disk wind should be related to the turbulent viscosity, we vary $\alpha$ and $C_{\rm w}$ independently to explore parameter space. The list of parameters for each model is summarized in Table~\ref{tbl:list}. 

\begin{table}
\caption{List of parameters of each model. The variable $\alpha$ indicates the strength of turbulent viscosity while $C_{\rm w}$ is a scaling factor for the disk wind.}
\label{tbl:list}
\centering
\begin{tabular}{l c c l}
\hline\hline
Model&  $\alpha$&                       $C_{\rm w}$&            Comment\\
\hline 
1&              $8 \times 10^{-3}$&     0&                              without disk wind\\
2&              $8 \times 10^{-3}$&     $2 \times 10^{-5}$&     weak magnetic field\\
3&              Eq.~(\ref{eq:alpha_evol})&      Eq.~(\ref{eq:cw_evol})&         strong magnetic field\\
4&              $2 \times 10^{-5}$&     $5 \times 10^{-7}$& described in Sec.~\ref{sec:parameters}\\
\hline
\end{tabular}
\end{table}

In model~1, the effect of the disk wind is not considered to clarify its effect on planet formation. The strength of the net vertical magnetic field is an important control parameter that determines $\alpha$ and $C_{\rm w}$ \citep{suzuki_etal10,okuzumi_hirose11}. If the net vertical magnetic field is sufficiently weak, $\alpha$ and $C_{\rm w}$ stay more or less constant during the disk evolution. Model~2 corresponds to this weak regime; we adopt $\alpha = 8 \times 10^{-3}$ and $C_{\rm w} = 2 \times 10^{-5}$ according to the MHD simulations of \citet{suzuki_etal10}. If the net vertical magnetic field is relatively strong,  $\alpha$ and $C_{\rm w}$ increase with time as gas materials are dispersed \citep{suzuki_etal10}. In model~3 we consider this strong regime, and we use
\begin{eqnarray}
\label{eq:alpha_evol}
\alpha = 8 \times 10^{-3} \times {\rm max}\left(1, \frac{0.01 \Sigma_{\rm g,ini}(r)}{\Sigma_{\rm g}(r)} \right),\\
\label{eq:cw_evol}
C_{\rm w} = 2 \times 10^{-5} \times {\rm max}\left(1, \frac{0.01 \Sigma_{\rm g,ini}(r)}{\Sigma_{\rm g}(r)} \right).
\end{eqnarray}
In addition, in model~4 we use $\alpha = 2 \times 10^{-5}$ and $C_{\rm w} = 5 \times 10^{-7}$. 

The parameter set for model~4 may require some explanation. The slope of gas surface density basically depends on $\alpha / C_{\rm w}$. In other words, if the ratio of $\alpha/C_{\rm w}$ is the same, the distribution of gas surface density takes the same form while the evolution timescale depends on $\alpha$ and $C_{\rm w}$. In model~4 we consider the case of a more efficient disk wind $(\alpha / C_{\rm w} = 40)$ than that of model~2 $(\alpha / C_{\rm w} = 400)$. The value of $\alpha$ is also changed because we wish to discuss the origin of the solar system's terrestrial planets in this paper. As we  show in Sects.~\ref{sec:results} and \ref{sec:discussion}, outward migration of 0.1 Earth-mass embryos is favorable for reproducing the observed properties of the solar system and $\alpha \sim 10^{-5}$ is required to avoid the saturation of the corotation torque for these bodies. Therefore, $\alpha = 2 \times 10^{-5}$ and $C_{\rm w} = \alpha / 40 = 5 \times 10^{-7}$ are adopted. While the evolution of the surface density of model~4 is slower than models~2 and ~3 because both $\alpha$ and $C_{\rm w}$ are smaller, the trend of the radial dependence of the surface density of this model represents a case more or less between model~2 and model~3. The initial amount of gas and the gas dispersal time are also changed in model~4. The scaling factor for the initial gas density $f_{\rm g}$ is increased by a factor of three. In the other models, $f_{\rm g}$ is set to unity. In addition, the disk would rapidly disperse $(\sim 0.1 {\rm Myr})$ after typical disk lifetimes of a few Myr, and additional effects are required to achieve this (e.g., photoevaporating winds) (e.g., \citealt{alexander_etal14}). In model~4, in order to mimic this disk evolution, $\alpha$ and $C_{\rm w}$ are increased by a factor of 100 after $t = 3 {\rm Myr}$.

For the initial distribution of embryos, we put 60 protoplanets with mass $M = 0.1 M_\oplus$ between $a=0.1-3$ AU. Their orbital separation, $\Delta$, measured in mutual Hill radii is set to 10. We perform three runs of simulations for each model with different initial locations of protoplanets.

For numerical integration, we use a fourth-order Hermite scheme \citep{makino_aarseth92}. When the physical radii of two bodies overlap, they are assumed to merge into one body, conserving total mass and momentum assuming perfect accretion. The physical radius of an embryo $R$ is determined by $R = (3M/4\pi \rho)^{1/3}$, where we assume the internal density of $\rho = 3 {\rm g~cm^{-3}}$.

\section{Results}
\label{sec:results}

\subsection{Disk evolution}
\begin{figure}
\resizebox{1.0 \hsize}{!}{\includegraphics{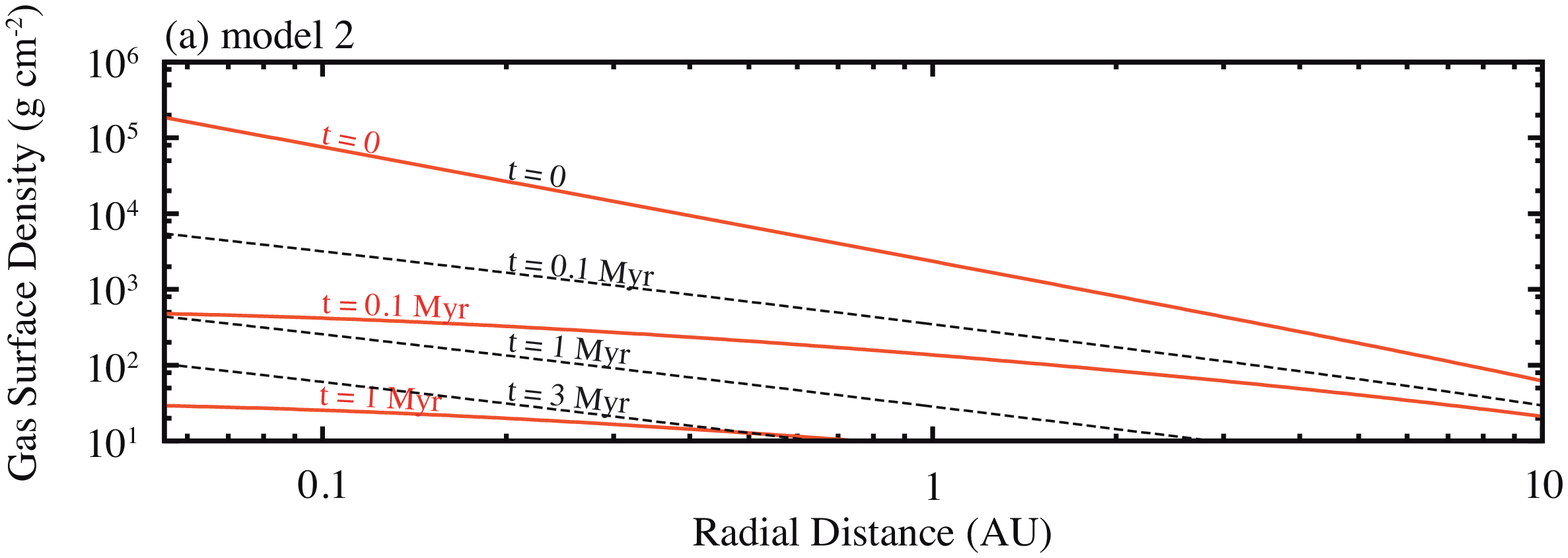}}
\resizebox{1.0 \hsize}{!}{\includegraphics{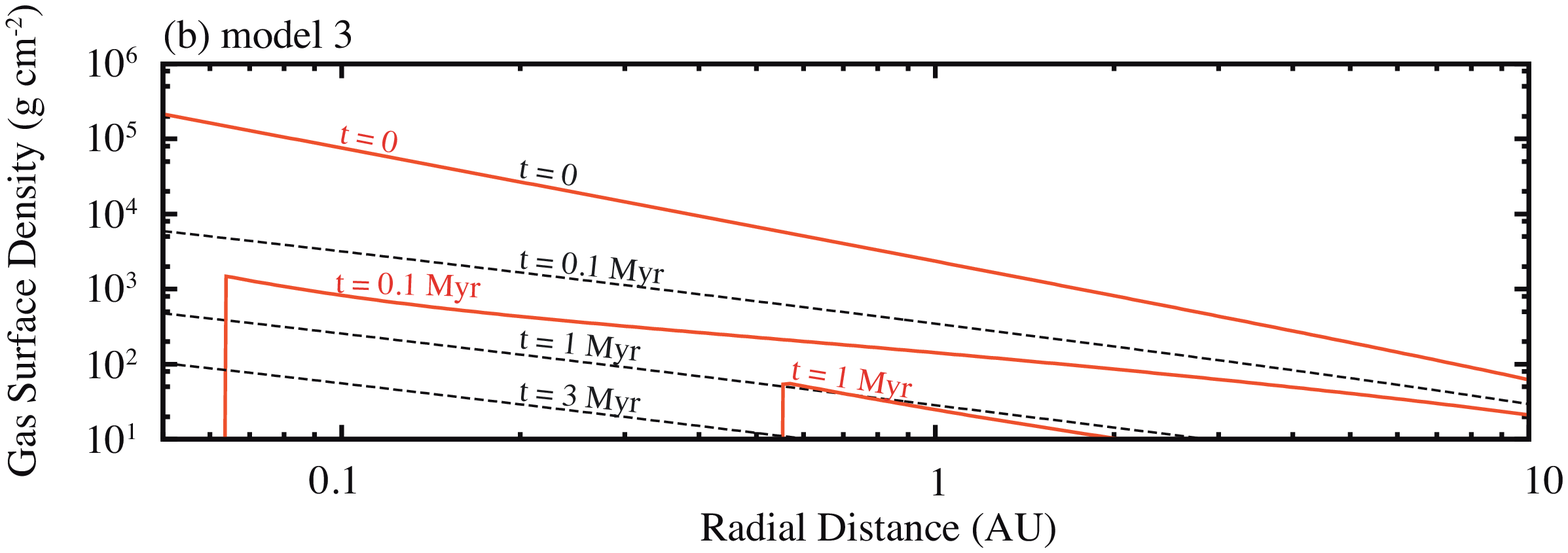}}
\resizebox{1.0 \hsize}{!}{\includegraphics{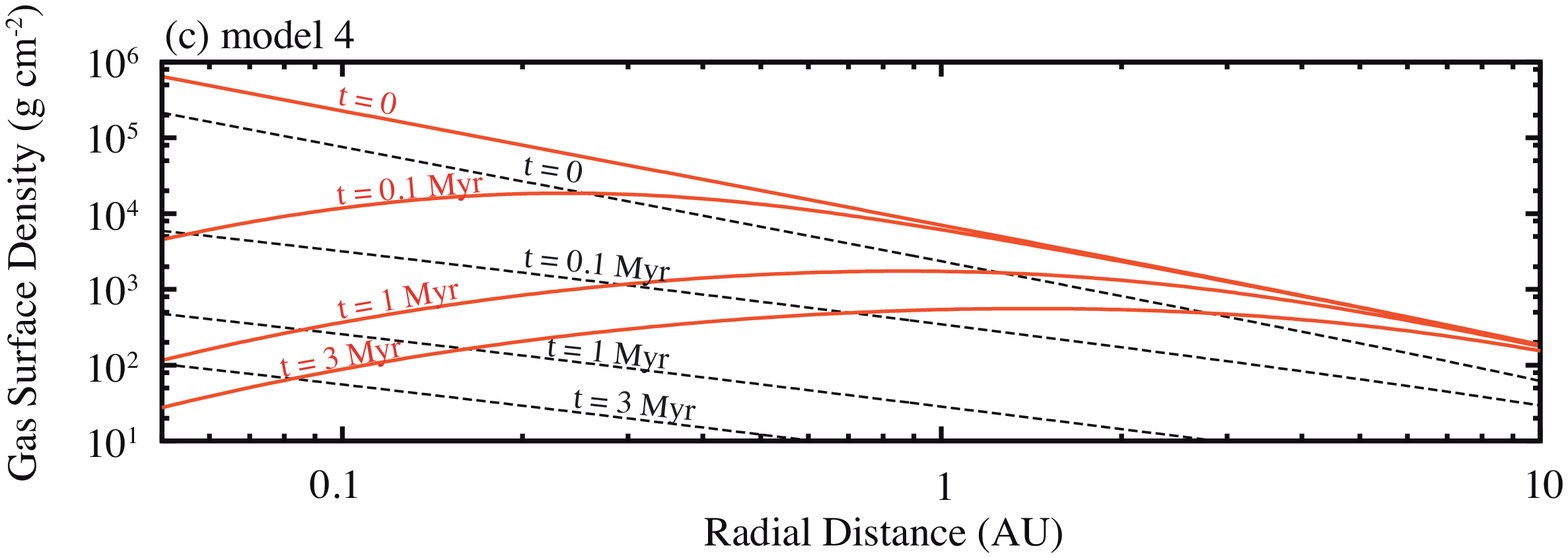}}
\caption{Evolution of gas surface density profile for (a) model~2, (b) model~3, and (c) model~4. Solid lines in each panel indicate the gas surface density at $t=0, 0.1 {\rm Myr}, 1 {\rm Myr}$, and $3 {\rm Myr}$, respectively. Dotted lines show results for model~1. In panels (a) and (b), the solid and dotted lines overlap for $t=0$ because of the initial conditions.
}
\label{fig:r-sigma}
\end{figure}

\begin{figure*}
\resizebox{1.0 \hsize}{!}{\includegraphics{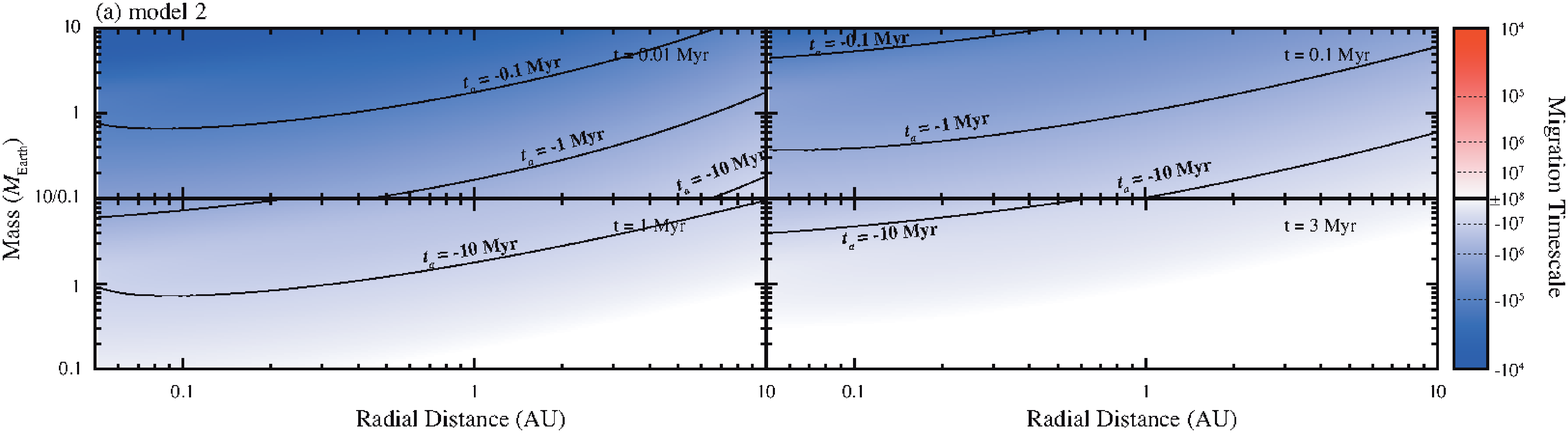}}
\resizebox{1.0 \hsize}{!}{\includegraphics{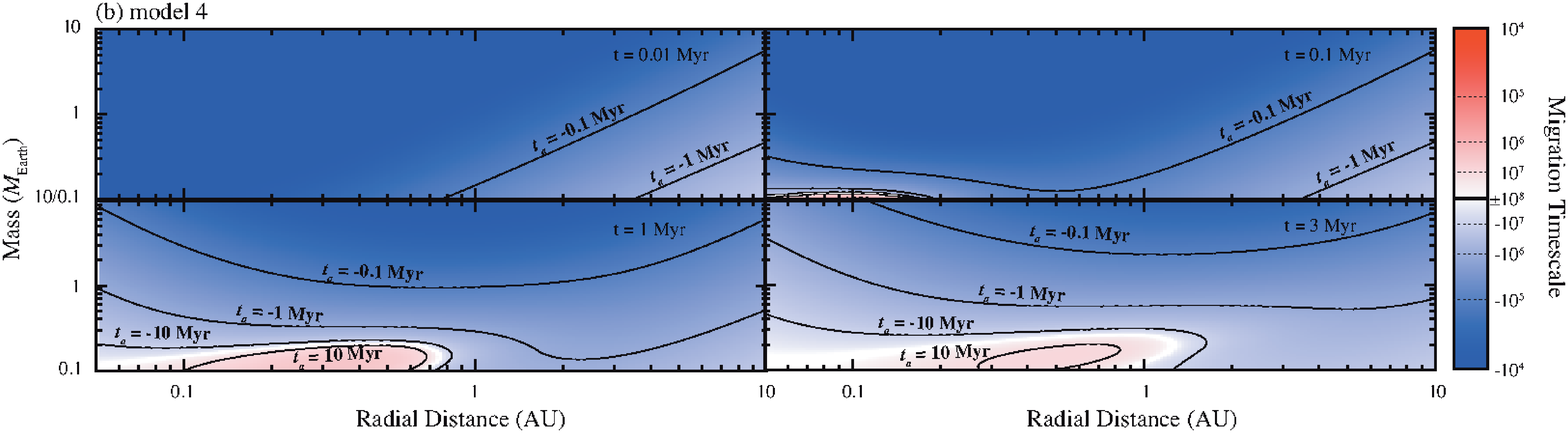}}
\caption{Time evolution of migration rate and direction for bodies with $e=0.01$ for (a) model~2 and (b) model~4. The color indicates the migration timescale. The contours show migration timescales of 0.1 Myr, 1 Myr, and 3 Myr. 
}
\label{fig:timemap}
\end{figure*}

We first show the disk evolution and its contribution to type I migration. The solid lines in Fig.~\ref{fig:r-sigma} show the evolution of the disk surface density for models~2-4 while the dotted lines in each panel represent the surface density for model~1. Figure~\ref{fig:timemap} shows the migration timescale $(t_a)$ for model~2 and model~4, where the migration rate and direction are indicated by the color (see color bar). When the migration timescale is negative, the direction of migration is inward. As we saw in Sect.~\ref{sec:model}, type I migration depends on the eccentricity of planets. We assume $e=0.01$ in these plots.

Figure~\ref{fig:r-sigma}(a) clearly shows that the disk evolution of model~2 is altered by the disk wind in the following two aspects. First, the slope of the disk becomes shallow in the inner region. Second, the disk is depleted faster than the case without  the disk wind. These two factors significantly reduce the inward migration of embryos. From the migration map of model~2 in Fig.~\ref{fig:timemap}(a), we find that the inward migration timescale for sub-Earth mass planets is over 1 Myr at $t=0.1 {\rm Myr}$, which means they no longer undergo migration. We also observe no migration at $t=1 {\rm Myr}$ and 3 Myr.

For the disk evolution of model~3 that has larger $C_{\rm w}$ and $\alpha$ shown in Fig.~\ref{fig:r-sigma}(b), an inner cavity is created inside a certain radius at which 99\% of the initial gas is dispersed according to Eqs. (\ref{eq:alpha_evol}) and (\ref{eq:cw_evol}). As time passes, the inner cavity grows gradually and the disk inner edge moves outward. The inside-out evacuation of the disk can significantly affect the orbital evolution of planets, which will be shown in Sect.~\ref{sec:orbital-evolution}.

We also examine the disk evolution for model~4, in which the radial profile of the surface density in the inner region is approximately between model~2 and model~3. In Figs.~\ref{fig:r-sigma} and \ref{fig:timemap} for model~2, we see the disk is depleted in the early stage of planet formation; therefore, a larger amount of initial gas disk is assumed $(f_{\rm g}=3)$. The evolution of the surface density in Fig.~\ref{fig:r-sigma}(c) shows that the slope becomes positive inside 1 AU. In order to get a positive surface density gradient inside 1 AU, we find that $\alpha / C_{\rm w} \lesssim 100$ is required. The disk lifetime is longer than model~2, which is more consistent with the observationally inferred disk lifetime. The migration map in Fig.~\ref{fig:timemap}(b) is more complicated. Because the effect of the disk wind is stronger  in the inner region, the orbital region where planets can move outward is limited to $r \lesssim 1 {\rm AU}$. The saturation of the corotation torque depends on the mass and the disk viscosity. If $\alpha = 2 \times 10^{-5}$ is assumed, the saturation effect is weakest for bodies with $M \simeq 0.1~M_\oplus$. Therefore, only sub-Earth-mass planets can undergo outward migration. It is also worth noting that planetary orbits can evolve until a few Myr because the disk is almost completely depleted after 10 Myr.

Before proceeding to the next subsection, where the results of \textit{N}-body simulations are shown,  here we briefly discuss how the inner disk profile is determined by viscous diffusion and the disk wind. We note that the disk wind brings about the variety in the radial dependence of the surface density in the inner region ranging from a clear hole (model~3) to mild inside-out evacuation (model~2). We suppose that the inner hole seen in model~3 is an extreme case. In a realistic situation low-density gas should remain in the inner region and a low level of  accretion would continue, rather than the formation of a clear inner hole, even at later times as seen in model~4, because the origin of the energy to drive the disk wind is the gravitational energy liberated by gas accretion \citep{suzuki_etal10}; the infall of a certain amount of gas is required to launch vertical outflows. \citet{suzuki_etal10} demonstrated the time evolution of a global disk by a simple model taking into account such energy-limited disk winds, which shows that the radial profile of the surface density is slightly modified at $\lesssim 1$ AU.

\subsection{Orbital evolution}
\label{sec:orbital-evolution}
\begin{figure*}
\resizebox{0.5 \hsize}{!}{\includegraphics{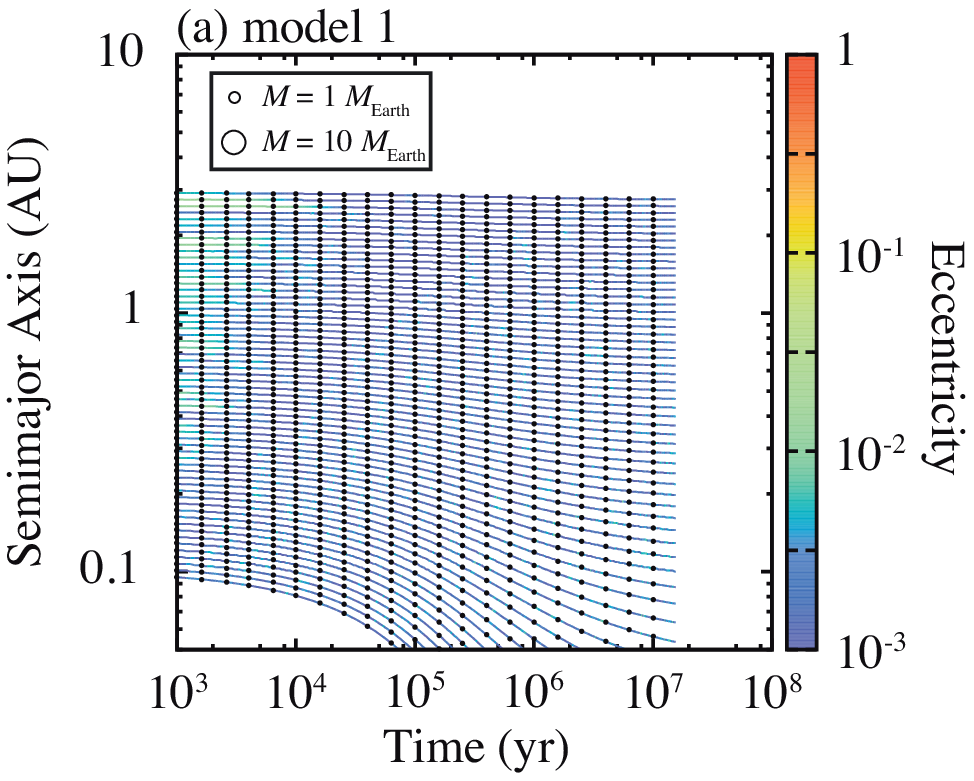}}
\resizebox{0.5 \hsize}{!}{\includegraphics{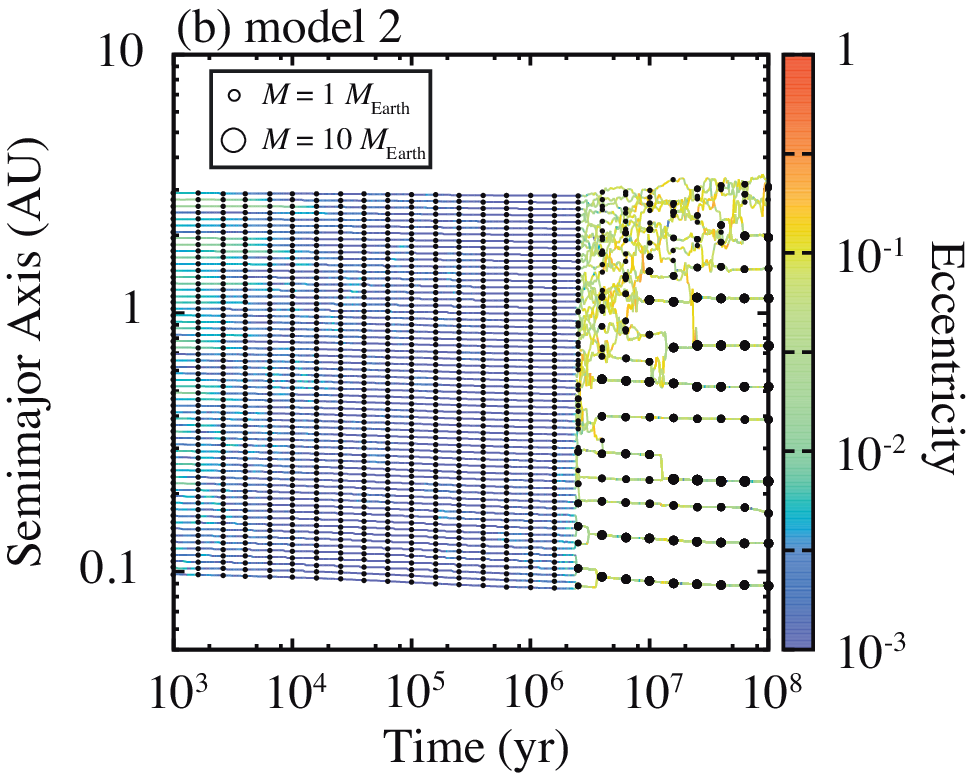}}
\resizebox{0.5 \hsize}{!}{\includegraphics{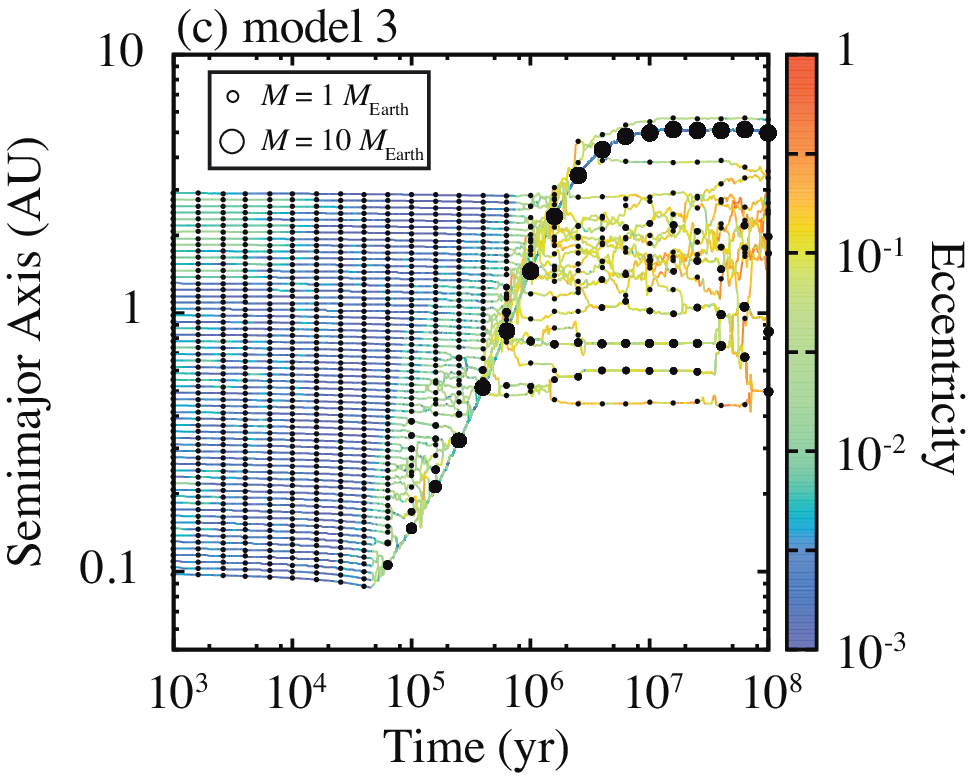}}
\resizebox{0.5 \hsize}{!}{\includegraphics{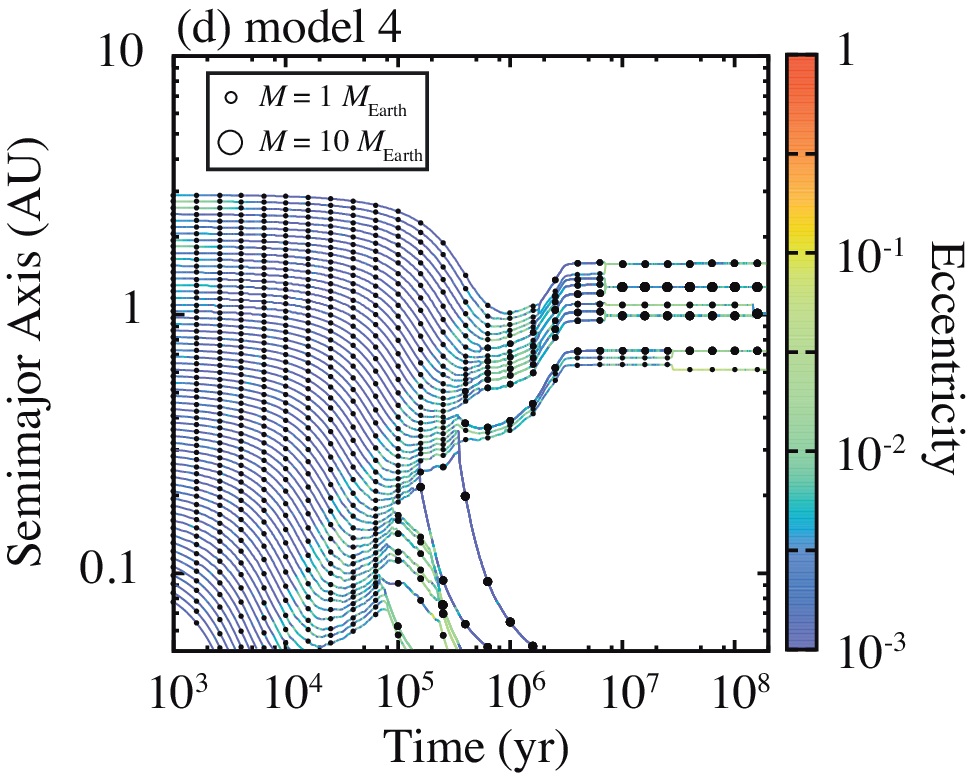}}
\caption{Orbital evolution of planetary embryos for (a) model~1, (b) model~2, (c) model~3, and (d) model~4. The filled circles connected with lines represent each body. The color of lines indicates the eccentricity (see color bar). The size of circles is proportional to the radius of the body.}
\label{fig:t-a}
\end{figure*}

\begin{figure}
\resizebox{1.0 \hsize}{!}{\includegraphics{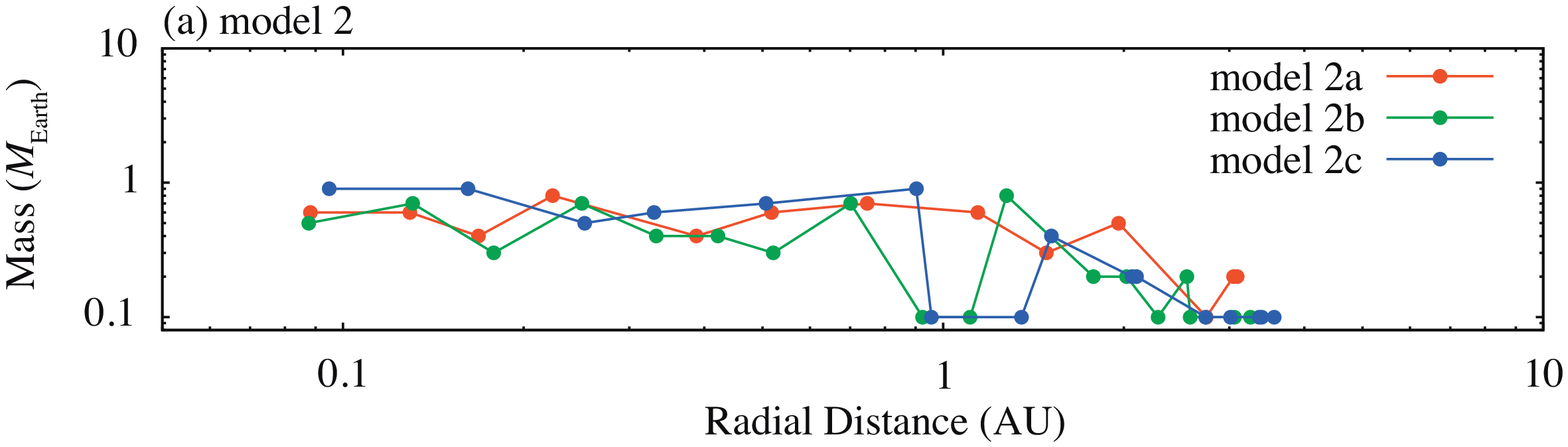}}
\resizebox{1.0 \hsize}{!}{\includegraphics{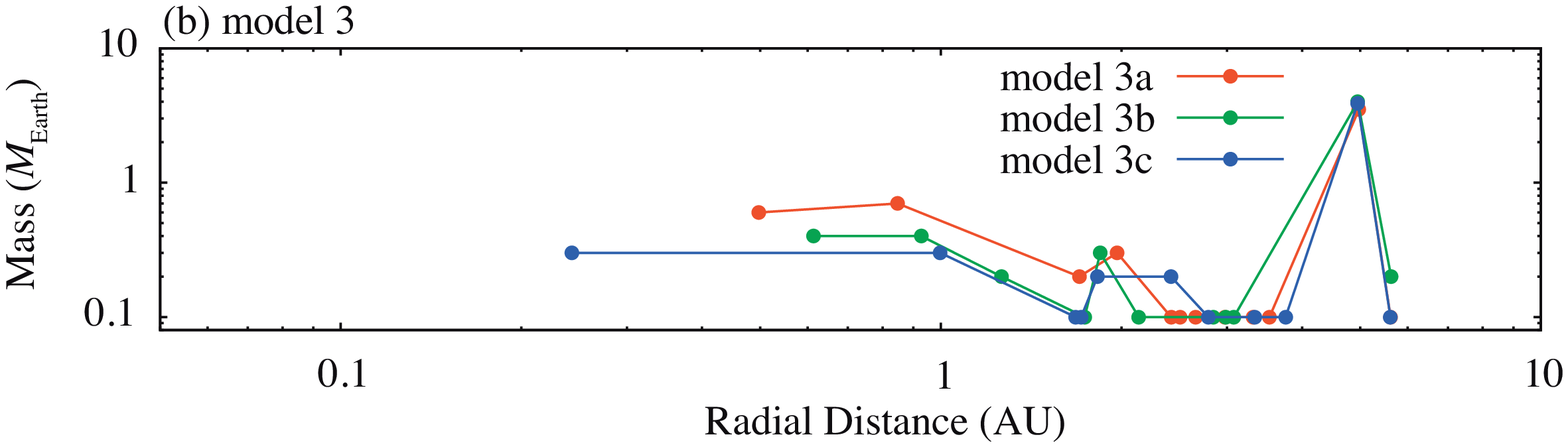}}
\resizebox{1.0 \hsize}{!}{\includegraphics{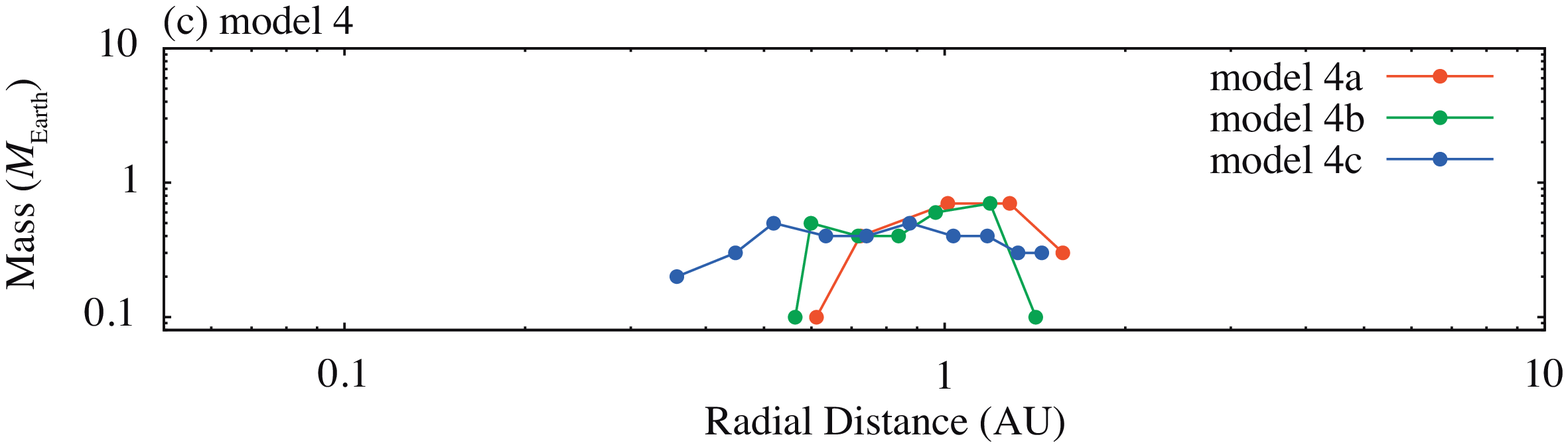}}
\caption{Final orbital configurations for (a) model~2, (b) model~3, and (c) model~4. Only for model~4a is the configuration at $t = 200 {\rm Myr}$  shown. Results of three different runs are plotted in each panel.}
\label{fig:a-m}
\end{figure}

We show the results of \textit{N}-body simulations in disks that evolved via viscous diffusion and a disk wind. Figure~\ref{fig:t-a} shows the orbital evolution of planets. The size of each filled circle represents the radii of bodies. The color of lines indicates the eccentricity of the planets. Each panel shows the results for model~1, model~2, model~3, and model~4. We performed three runs for each model, and typical runs are shown in the plots. Figure~\ref{fig:a-m} shows the final orbital configurations in the semimajor axis-mass plane for all runs of model~2, model~3, and model~4 (the results for model~1 are not shown), where planets that formed through the same run are connected with the same line.

In Fig.~\ref{fig:t-a}(a), the results for model~1 (in which the disk wind is not considered) is shown for comparison. Planetary embryos migrate inward and 12 bodies are lost by $t= 20 {\rm Myr}$. The computational cost is huge for this run because the bodies do not collide with each other and the orbital period of the innermost planet is short; therefore, the computation was stopped at $t = 20 {\rm Myr}$, which took over four months of CPU time. Orbit crossings are not observed before $t = 20 {\rm Myr}$; however, the system is expected to be destabilized at $\sim 100$ Myr according to previous studies on the stability of multi-planet systems \citep{chambers_etal96}.

In the results for model~2, we find that planetary embryos do not migrate inward any more as expected from Figs.~\ref{fig:r-sigma} and \ref{fig:timemap}. In this model, as the disk dissipates, the eccentricity of the embryos is excited and they exhibit orbit crossings, leading to giant impacts between the bodies. At $t = 100 {\rm Myr}$, the number of planets has decreased to 13 and their eccentricities are between about 0.01 and 0.1. The final orbital configurations are almost the same among the three different runs (see Fig.~\ref{fig:a-m}(a)). The averaged mass of the largest body in each run is $\simeq 0.9 M_\oplus$. The typical orbital separation between the planets is $\simeq 10-40 r_{\rm H}$.

For model~3, the orbital evolution is significantly different from that without a disk wind. As seen in Fig.~\ref{fig:r-sigma}(b), the disk's inner edge moves outward. The migrating inner edge sweeps up embryos and almost all bodies migrate outward. As a result, no planets are left inside 0.4 AU and a large planet forms at 5 AU. The eccentricities of the planets are $\sim 0.04-0.3$ (some bodies are still exhibiting orbit crossings). In the other two runs, the results are almost the same (see Fig.~\ref{fig:a-m}(b)). The mass of the largest body is $\simeq 3.8 M_\oplus$, while the second largest planet is located at around $1 {\rm AU}$ with a mass of $\simeq 0.5 M_\oplus$.

We observe an interesting orbital evolution in model~3 in which most protoplanets are swept up by the disk edge before $t \simeq 0.3 {\rm Myr}$ but afterward several bodies drift through the edge. We briefly discuss how bodies follow the migration of the disk edge or not. First, just as there are for torques on planets, there are two different positive torques, namely the type I migration torque and the edge torque. As seen in Sect.~\ref{sec:model}, the former can become positive owing to the corotation torque, which depends on the slope of the gas surface density. The latter is a torque recently found by \citet{ogihara_etal10}. When a body with non-zero  eccentricity straddles a sharp disk inner edge, the body gains a net positive torque from the disk depending on the sharpness of the edge and on the eccentricity. Whether planets can follow the movement of the disk edge depends on the migration speeds of the planets and the edge. If the outward migration rate of the planets due to the positive torque (which is the sum of the type I migration torque and the edge torque) is smaller than the migration rate of the edge, the planets go through the disk edge and stay in place. Through a series of test calculations, we find that the positive torque exerted on the planets is usually large for this model and 0.1 Earth-mass planets can follow the disk edge. Therefore, there is another reason why some planets drift through the disk, which is that small embryos are scattered inside the disk edge by larger bodies. Larger bodies grow at the disk edge by sweeping up smaller embryos. At $t \simeq 1 {\rm Myr}$, the planets at the edge already have masses of $\simeq 3.5 M_\oplus$ and scatter embryos inside the edge. 

The results for model~4 are also interesting. Before $t \simeq 0.02$ Myr, all bodies undergo inward migration, but afterward  $\sim 0.1$ Earth-mass embryos in close-in orbits move outward while bodies in distant regions migrate inward, leading to a convergence of planets at around 1 AU. For $\alpha = 2 \times 10^{-5}$, the positive corotation torque can be saturated for bodies larger than $\simeq 0.3$ Earth masses; as a result, several planets that undergo collisional growth migrate inward and are lost to the central star. As the disk disperses after $t = 3 {\rm Myr}$, all the bodies cease their migration. At $t = 100 {\rm Myr}$, a system with six planets with a maximum mass of $0.7 M_\oplus$  and $e \simeq 0.01$ had formed. The orbital separations between the planets are relatively small (down to $\lesssim 10 r_{\rm H}$), and giant impacts can occur after $t = 100 {\rm Myr}$. Thus we continued simulation until $t = 200 {\rm Myr}$ for this run. The system underwent a giant impact at $t = 145 {\rm Myr}$, which is discussed in Sect.~\ref{sec:discussion}. In the other two runs, although the mass distributions are slightly different from each other, the results are qualitatively the same (Fig.~\ref{fig:a-m}(c)).

\section{Origin of the solar system's terrestrial planets}
\label{sec:discussion}
Here we discuss our results in terms of the origin of terrestrial planets in the solar system.  The important characteristics are the radial mass concentration and small eccentricities of planets, which we characterize using two statistical measures, $S_{\rm c}$ and $S_{\rm d}$.
The radial mass concentration statistic is \citep{chambers_01}
\begin{eqnarray}
\label{eq:sc}
S_{\rm c} =  \max\left( \frac{\sum m_j}{\sum m_j [\log_{10}(a/a_j)]^2} \right),
\end{eqnarray}
where $j=1,2,...,N$. The angular momentum deficit, which represents the deviation from circular and coplanar orbits, is \citep{laskar_97}
\begin{eqnarray}
\label{eq:sd}
S_{\rm d} = \frac{\sum m_j \sqrt{a_j} \left[ 1 - \cos(i_j) \sqrt{1-e_j^2}\right]}{\sum m_j \sqrt{a_j}},
\end{eqnarray}
where $i_j$ is the orbital inclination of the body $j$. 

The radial mass concentration of the solar system's terrestrial planets is $S_{\rm c} = 89.9$, which indicates high radial mass concentration  because the masses of Mercury and Mars are small and the orbital distance between Venus and the Earth is also relatively small. Previous studies have shown that it is difficult to reproduce such a high mass concentration (e.g., \citealt{raymond_etal09}; \citealt{morishima_etal10}); a way to account for this statistic in \textit{N}-body simulations is to assume a mass distribution initially confined to a narrow annulus (\citealt{morishima_etal08}; \citealt{hansen_09}), or in other words to assume an initially large $S_{\rm c}$. In the grand tack model (e.g., \citealt{walsh_etal11}; \citealt{jacobson_morbidelli14}), Jupiter sweeps up to $\sim 1.5 {\rm AU}$ and embryos are eventually confined to a relatively compact region. However, they assumed an inner cutoff at $\simeq 0.7 {\rm AU}$, which means that $S_{\rm c}$ is already relatively large in the initial state.

The angular momentum deficit for the terrestrial planets in the solar system is $S_{\rm d} = 0.0018$, which means that the eccentricities of the terrestrial planets in the solar system are relatively small ($e \simeq 0.01 - 0.1$). We note that Jupiter and Saturn increase $S_{\rm d}$ after their formation, which is ignored in this paper. Therefore, the necessary condition for our results to explain terrestrial planets in the solar system is that $S_{\rm d} \lesssim 0.0018.$

\begin{table}
\caption{Statistical measures for final planetary systems. $S_{\rm c}$ is a measure of the radial mass concentration. $S_{\rm d}$ measures how a system deviates from circular and coplanar orbits. The last column shows the timing of the last giant impact, although last impact events do not occur within the simulation time for model~1 and model~4 (except for 4a).}
\label{tbl:discussion}
\centering
\begin{tabular}{l c c c}
\hline\hline
Run&    $S_{\rm c}/89.9$&       $S_{\rm d}/0.0018$&             $t_{\rm imp, last}$ (Myr)\\
\hline 
1a&             0.042&                  0.0016&                         $> 20$\\
1b&             0.050&                  0.0016&                         $> 20$\\
1c&             0.050&                  0.0013&                         $> 20$\\
2a&             0.048&                  1.7&                                    61\\
2b&             0.052&                  1.5&                                    30\\
2c&             0.047&                  9.1&                                    80\\
3a&             0.086&                  3.7&                                    99\\
3b&             0.12&                   1.5&                                    57\\
3c&             0.098&                  0.87&                           72\\
4a&             0.80&                   0.046&                          $145$\\
4b&             0.81&                   0.0045&                         $> 100$\\
4c&             0.35&                   0.020&                          $> 100$\\
\hline
\end{tabular}
\end{table}

Table~\ref{tbl:discussion} shows the statistical measures for all runs. In the cases of model~1 (1a, 1b, 1c) and model~2 (2a, 2b, 2c), $S_{\rm c}$ is kept small from the beginning of the simulations because the planets do not undergo convergent migration. The initial value of $S_{\rm c,0}$ is 5.4. In the results of model~3 (3a, 3b, 3c), $S_{\rm c}$ is slightly increased to $\simeq 9$. The final system is more concentrated than the initial distribution, but the value is still not consistent with the value of the current solar system. In the results of model~4 (4a, 4b, 4c), $S_{\rm c}$ is as large as the value of the inner solar system ($S_{\rm c}$ is between 30 and 70). The mass concentration at around 1 AU is clearly seen in Fig.~\ref{fig:t-a}(d).

Regarding the angular momentum deficit, $S_{\rm d}$ for model~1 is small because planets do not exhibit orbit crossings. In contrast, in model~2 and model~3, planets experience orbit crossings and the gas disk is depleted early, thus $S_{\rm d}$ is larger than that of model~1. We neglect the forced eccentricity by Jupiter and Saturn such that additional damping may be required for model~2 and model~3. In the results of model~4, $S_{\rm d}$ is sufficiently small $(\simeq 4 \times 10^{-5})$. 

The timing of a moon-forming impact is also a key constraint on formation models of the solar system. According to recent work, the time of the last giant impact on Earth could have been significantly late (50-150 Myr after the formation of the first solids in the solar system) (e.g., \citealt{touboul_etal07}; \citealt{allegre_etal08}; \citealt{jacobson_etal14}). The timing of the latest impact for model~2 is $t \sim 50$ Myr, and that for model~3 is $t \simeq 50-100$ Myr, consistent with the timing of the last giant impact on Earth. Particularly interesting is that the timing of the last giant impacts is delayed after 100 Myr for model~4. In fact, we continue simulation for 4a until $t = 200 {\rm Myr}$ and find that the last giant impact occurs at $t = 145 {\rm Myr}$. 

In conclusion, we find that a disk wind can increase the radial mass concentration. In particular, $S_{\rm c}$ can be as large as that of the solar system even if we assume that the initial embryo distribution is extended to 0.1 AU. In the result of 4a, the orbital properties including the timing of the moon-forming impact are similar to those of the  solar system. We have only performed simulations in a limited number of cases; however, we can identify a significantly large parameter space that may allow us to find a best-fit model for the observed properties of the solar system. To assess the possibility, we have to explore a broad range of parameter space to find a condition that can create a solar system analog.

We also discuss one other problem in the formation of the solar system. Recent numerical simulations on the evolution of the snow line in the solar system that solve the detailed radiative energy transfer have shown that the snow line moves in the disk and comes inside Earth's orbit to $\simeq 0.5 {\rm AU}$ (e.g., \citealt{oka_nakamoto11}; \citealt{davis05}; \citealt{garaud_lin07}) during a phase with a disk accretion rate of $\sim 10^{-10} M_\odot {\rm yr}^{-1}$. If planetesimals formed during this stage, the terrestrial planets in the solar system should have formed from icy planetesimals and contain significant amounts of water, which is not consistent with the current water content of the terrestrial planets. We can provide a pathway to avoid this problem. Protoplanets that formed from rocky planetesimals inside the snow line within $\lesssim 0.5 {\rm AU}$ migrate outward in disks that are affected by a disk wind, resulting in water-poor planets at around 1 AU.

\section{Conclusions}
\label{sec:conclusions}
We performed \textit{N}-body simulations that include the effects of a disk wind and found that the orbital evolution of terrestrial planets and their precursor protoplanets can significantly differ from the results of previous simulations. Because of the disk wind, the efficiency of which depends approximately on $r^{3/2}$, gas materials blow out from the surface of the disk from the inside out, leading to a shallower surface density profile in the inner region $(\lesssim 1 {\rm AU})$. Planets with masses less than a few Earth masses would no longer migrate inward. The orbital evolution depends on the parameters ($\alpha$, $C_{\rm w}$), and in fact, 0.1 Earth-mass embryos inside 1 AU never undergo significant inward migration for all the parameters in our simulations. In an extreme case in which the vertical magnetic field is strong, we observed the  inner edge of the disk move outward and sweep up embryos.

In a certain parameter range, we found that planetary embryos move outward. When $\alpha / C_{\rm w}$ is less than $\sim 100$, the surface density slope becomes positive in the inner region, which can induce convergent migration to around 1 AU. In order for outward migration to take place, the corotation torque should not be saturated, which depends on diffusion. For 0.1 Earth-mass embryos, the saturation is inhibited when $\alpha$ is $\simeq 10^{-5}$. Thus, we performed additional simulations with $\alpha = 2 \times 10^{-5}$ and $C_{\rm w} = 5 \times 10^{-7}$ and found that the resultant systems are radially concentrated at around 1 AU. This is an important result. In this case, the observed constraints of the inner solar system (e.g., eccentricity, radial mass concentration, late moon-forming impact) may be reproduced. This expectation requires that the turbulent viscosity be quite low $(\alpha \sim 10^{-5})$ in the terrestrial planet-forming region in the protosolar nebula. Although we focus on the properties of the terrestrial planets, other observed properties in the solar system (e.g., the dynamically excited asteroid belt) should be reproduced in future work. A combination of our model with the grand tack model \citep{walsh_etal11}, which explains asteroid excitation, may be a possible scenario. In addition, a disk wind can play other important roles. For example, if the snow line had been located inside 1 AU, the Earth would have accreted a significant amount of icy planetesimals. However, this problem can be overcome if the Earth formed from rocky protoplanets that migrated outward owing to the effect of the disk wind.

The disk wind can also apply to exoplanet formation models. A large number of close-in low-mass planets have been discovered, and thus we can statistically compare observational results with theoretical models. To do so, we need to explore parameter space $(\alpha, C_{\rm W})$, which we leave for future work. If disk winds operate efficiently in exoplanet systems, super-Earths would tend to pile up at around 1 AU rather than 0.1 AU.

\begin{acknowledgements}
We thank the anonymous referee for helpful comments. M.O. thanks Shigeru Ida and Alessandro Morbidelli for helpful discussions. We also thank Jennifer M. Stone for valuable comments. Numerical computations were in part conducted on the general-purpose PC farm at the Center for Computational Astrophysics, CfCA, of the National Astronomical Observatory of Japan. H.K. gratefully acknowledges support from Grant-in-Aid for Scientific Research (B) (26287101). S.I. is supported by Grant-in-Aid for Scientific Research (23244027, 23103005).
\end{acknowledgements}

{}



\end{document}